\documentclass[preprint,aps,prd,showpacs,nofootinbib]{revtex4}
\usepackage{mathrsfs}
\usepackage{amsmath}
\usepackage{graphicx}
\usepackage{subfigure}

\newcommand{\bea}{\begin{eqnarray}}
\newcommand{\eea}{\end{eqnarray}}
\newcommand{\beq}{\begin{equation}}
\newcommand{\eeq}{\end{equation}}

\def\/{\over}

\begin{document}

\title{Lamb Shift for static atoms outside a Schwarzschild black hole }
\author{  Wenting Zhou$^{1}$ and Hongwei Yu$^{1,2}$\footnote{Corresponding author}  }
\affiliation{$^1$ Department of Physics and Key Laboratory of Low
Dimensional Quantum Structures and Quantum
Control of Ministry of Education,\\
Hunan Normal University, Changsha, Hunan 410081, China \\
$^2$ Center for Nonlinear Science and Department of Physics, Ningbo
University,  Ningbo, Zhejiang 315211, China}

\begin{abstract}
We study,  by separately calculating the contributions of vacuum
fluctuations and radiation reaction to the atomic energy level
shift, the Lamb shift of a static two-level atom interacting with
real massless  scalar fields in the Boulware, Unruh and
Hartle-Hawking vacuums outside a Schwarzschild black hole.  We find
that in the Boulware vacuum, the Lamb shift gets a correction
arising as a result of the backscattering of vacuum field modes off
the space-time curvature, which is reminiscent of the correction to
the Lamb shift induced by the presence of cavities.  However, when
the Unruh and Hartle-Hawking vacua are concerned, our results show
that the Lamb shift behaves as if the atom  were irradiated by a
thermal radiation or immersed in a thermal bath at the Hawking
temperature, depending on whether the scalar field is in the Unruh
or the Hartle-Hawking vacuum. Remarkably, the thermal radiation is
always backscattered by the space-time geometry.

\end{abstract}
 \maketitle

\baselineskip=16pt

\section{Introduction}

The Lamb shift, the shift of the energy levels of an atom that
arises as a result of the coupling of the atom to vacuum
fluctuations of quantum fields in a flat space-time, is one of the
most remarkable observable effects predicted by quantum field
theory.  Later, it has been shown that this radiative energy level
shift of an atom can be modified by the presence of
cavities~\cite{Meschede90} and the noninertial motion of the atom
itself~\cite{Audretsch95,Passante98,L.Rizzuto07,ZhuYu10}, which
alter the vacuum fluctuations and yield a Lamb shift different from
its original value.

Here, we are concerned with the Lamb shift of a static atom in the
exterior of a spherically symmetric black hole interacting with
vacuum fluctuations of quantum fields. When a curved space-time is
considered as opposed to a flat one, a delicate issue then arises as
to how the vacuum state of the quantum fields is determined.
Normally, a vacuum state is associated with nonoccupation of
positive frequency modes. However, the positive frequency of field
modes is defined with respect to the time coordinate. Therefore, to
define positive frequency, one has to first specify a definition of
time. In a spherically symmetric black hole background, one
definition is the Schwarzschild time, $t$, and it is a natural
definition of time in the exterior region. The vacuum state, defined
by requiring normal modes to be positive frequency with respect to
the Killing vector $\partial/
\partial t$ with respect to which the exterior region is static, is
called the Boulware vacuum. Other possibilities that have been
proposed are the Unruh vacuum~\cite{Unruh} and the Hartle-Hawking
vacuum~\cite{Hartle-Hawking}. The Unruh vacuum is defined by taking
modes that are incoming from $\mathscr{J}^-$ to be positive
frequency with respect to $\partial/ \partial t$, while those that
emanate from the past horizon are taken to be positive frequency
with respect to the Kruskal coordinate $\bar u$, the canonical
affine parameter on the past horizon. The Hartle-Hawking vacuum, on
the other hand, is defined by taking the incoming modes to be
positive frequency with respect to $\bar v$, the canonical affine
parameter on the future horizon, and outgoing modes to be positive
frequency with respect to $\bar u$.  The calculations of the values
of physical observables, such as the expectation values of the
energy-momentum tensor and the response rate of an Unruh detector in
these vacuum states, have yielded the following  physical
understanding: (i) The Boulware vacuum corresponds to our familiar
concept of a vacuum state at large radii. It would be the vacuum
state outside a massive spherical body of radius only slight larger
than its Schwarzschild radius; (ii) the Unruh vacuum is the vacuum
state that best approximates the state that we would obtain
following the gravitational collapse of a massive body to a black
hole; (iii) the Hartle-Hawking state, however, does not correspond
to our usual notion of a vacuum, but describes a black hole in
equilibrium with a sea of thermal radiation.

In the current paper, we will study the Lamb shift of a static
two-level atom outside a four-dimensional Schwarzschild black hole
in interaction with massless quantum scalar fields in all the above
three vacuum states. It should be pointed out that a fully realistic
treatment of the Lamb shift requires considering a multilevel atom
coupled to the electromagnetic field. However, the purpose of the
present paper is to bring to light the essential features of the
full problem in a simple model and keep the discussion as clear and
transparent as possible. Our calculations of the Lamb shift make use
of an elegant formalism suggested by Dalibard, Dupont-Roc, and
Cohen-Tannoudji(DDC)~\cite{Dalibard82,Dalibard84} which allows a
separation of the contributions of vacuum fluctuations and the
radiation reaction to the energy shifts. This separation is also
interesting from a conceptual point of view since the Lamb shift is
usually associated with vacuum fluctuations alone. In  previous
studies~\cite{hyprd07}, we have separately calculated the
contributions of vacuum fluctuations and the radiation  reaction to
the spontaneous excitation rate of a static atom outside a
Schwarzschild black hole, and find that the atom would spontaneously
excites as if it were irradiated by or immersed in a thermal
radiation at the Hawking temperature.

\section{General formalism}

Let us consider a two-level atom in interaction with a quantum real
massless scalar field outside a Schwarzschild black hole. The metric
of the space-time can be written in terms of the Schwarzschild
coordinates as
 \beq
 ds^2= \bigg(1-{2M\over r}\bigg)\;dt^2-\bigg(1-{2M\over
r}\bigg)^{-1}\;dr^2-r^2\,(d\theta^2+\sin^2\theta\,d\varphi^2)\;,
 \eeq
 where $M$ is the mass of the black hole. Without loss of
generality, we assume a pointlike two-level atom on a stationary
space-time trajectory $x(\tau)$, where $\tau$ denotes the proper
time on the trajectory. The stationarity of the trajectory
guarantees the existence of stationary atomic states, $|+ \rangle$
and $|- \rangle$, with energies $\pm{1\/2}\omega_0$ and a level
spacing $\omega_0$. The Hamiltonian that describes the time
evolution of the atom-field interacting system can be given by
\cite{Audretsch94,Audretsch95}
 \beq
H(\tau)=H_A(\tau)+H_F(\tau)+H_I(\tau)\;,\label{H}
 \eeq
 where
 \beq
H_A(\tau)=\hbar\omega_0S_z(\tau)\;,
 \eeq
 \beq
H_F(\tau)=\int d^3 k \,\hbar \omega_{\vec{k}} a_{\vec{k}}^\dag
a_{\vec{k }}{dt\/d \tau}\;,
 \eeq
 \beq
H_I(\tau)={\mu}S_2(\tau)\phi(x(\tau))\;.\label{HI}
 \eeq
 Here $a_{\vec{k}}^\dag$ and $a_{\vec{k }}$ are the creation and
annihilation operators for a scalar particle with  momentum
$\vec{k}$, and $\mu$ is a coupling constant which we assume to be
small. $S_z$, $S_+$, and $S_-$ are the pseudospin operators of the
atom, $S_z(0)=(1/2)(|+\rangle\langle+|-|-\rangle\langle-|)$ and
$S_2(0)=(i/2)(S_--S_+)$. Here $S_+(0)=|+\rangle\langle-|$, and
$S_-(0)=|-\rangle\langle+|$. The field operators $\phi$ is evaluated
along the trajectory $x(\tau)$ of the atom.

Then we can write down the Heisenberg equations of motion for the
dynamical variables of the atom and field from the Hamiltonian of
the system in Eq.~(\ref{H}). The solutions of the equations of
motion can be split into the two parts: a free part, which is
present even in the absence of the coupling, and a source part,
which is caused by the interaction of the atom and field. We assume
that the initial state of the atom is $|b\rangle$ and the scalar
field is in a vacuum state $|0\rangle$. To identify the two distinct
contributions of vacuum fluctuations and radiation reaction to the
energy level shift of a two-level atom, we use DDC's formalism to
choose a symmetric ordering between the atom and field variables,
and separate the vacuum fluctuations and radiation reaction
contributions to the rate of change of an arbitrary atomic
observable $G(\tau)$. Following the procedures that have been shown
in Refs.~\cite{Dalibard84,Audretsch95}, we take the average over the
vacuum state of the scalar field, and obtain
 \beq
 \bigg\langle
0\bigg|\bigg({d G(\tau)\/d\tau}\bigg)_{vf,rr}\bigg|
0\bigg\rangle={i\/\hbar}[H^{eff}_{vf,rr}(\tau),G(\tau)]+\text{non-Hamiltonian\
terms}\;,
 \eeq
 where the non-Hamiltonian parts are related to
relaxation effects, and in order $\mu^2$, the effective Hamiltonians
can be expressed as
 \bea
H^{eff}_{vf}(\tau)&=&{i\mu^2\/2\hbar}\int^\tau_{\tau_0}d\tau'
C^F(x(\tau),x(\tau'))[S_2^f(\tau),S_2^f(\tau')]\;,\label{Htf}\\
H^{eff}_{rr}(\tau)&=&-{i\mu^2\/2\hbar}\int^\tau_{\tau_0}d\tau'
\chi^F(x(\tau),x(\tau'))\{S_2^f(\tau),S_2^f(\tau')\}\;.\label{Hrr}
 \eea
 Here $[\ ,\ ]$ and $\{\ ,\ \}$ are the commutator and
anti-commutator respectively, and the subscripts `` $vf$ " stands
for vacuum fluctuations and ``$rr$" for radiation reaction. The
statistical functions of the field  $C^F$ and $\chi^F$ are also
called symmetric correlation function and linear susceptibility of
the field. They are defined as
 \bea
 C^{F}(x(\tau),x(\tau')) &=& {1\/2}{\langle} 0| \{ \phi^f
(x(\tau)), \phi^f(x(\tau')) \} | 0 \rangle\;, \label{general form of
Cf}\\
\chi^F(x(\tau),x(\tau')) &=& {1\/2}{\langle} 0| [
\phi^f(x(\tau)),\phi^f (x(\tau'))] | 0 \rangle\;. \label{general
form of Xf}
 \eea
Taking the expectation value of Eqs.~(\ref{Htf}) and (\ref{Hrr}) in
the atom's initial state $|b\rangle$, we can obtain the radiative
energy shifts of the atom's level $|b\rangle$ due to the vacuum
fluctuations and radiation reaction,
 \bea
(\delta E_b)_{vf}&=&-{i\mu^2\/\hbar}\int^\tau_{\tau_0}d\tau'
C^F(x(\tau),x(\tau'))(\chi^A)_b(\tau,\tau')\;,\label{Evf}\\
(\delta E_b)_{rr}&=&-{i\mu^2\/\hbar}\int^\tau_{\tau_0}d\tau'
\chi^F(x(\tau),x(\tau'))(C^A)_b(\tau,\tau')\;,\label{Err}
 \eea
where $(C^A)_b$ and $(\chi^A)_b$, the symmetric correlation function
and the linear susceptibility of the atom, are defined as
 \bea
(C^A)_b(\tau,\tau')&=&{1\/2}\langle
b|\{S_2^f(\tau),S_2^f(\tau')\}|b\rangle\;,\\
(\chi^A)_b(\tau,\tau')&=&{1\/2}\langle
b|[S_2^f(\tau),S_2^f(\tau')]|b\rangle\;,
 \eea
 which are
characterized by the atom itself. Explicitly, the statistical
functions of the atom can be given as
 \bea
(C^A)_{b}(\tau,\tau')&=&\frac{1}{2}\sum_{d}|\langle
b|S_2(0)|d\rangle|^2
   (e^{i\omega_{bd}\Delta\tau}+e^{-i\omega_{bd}\Delta\tau})\;,\\
(\chi^A)_{b}(\tau,\tau')&=&\frac{1}{2}\sum_{d}|\langle
b|S_2(0)|d\rangle|^2
   (e^{i\omega_{bd}\Delta\tau}-e^{-i\omega_{bd}\Delta\tau})\;,
 \eea
where $\omega_{bd}=\omega_b-\omega_d$, $\Delta \tau=\tau-\tau'$
and the sum extends over a complete set of atomic states.

\section{Lamb shifts of static atoms outside a black hole.}

In the exterior region of the Schwarzschild black hole, a complete
set of normalized basis functions for the massless scalar field that
satisfy the Klein-Gordon equation is given by
 \bea
 \overrightarrow{u}_{\omega
lm}=(4\pi\omega)^{-\frac{1}{2}}e^{-i\omega
t}\overrightarrow{R}_l(\omega|r)Y_{lm}(\theta,\varphi)\;,\\
\overleftarrow{u}_{\omega lm}=(4\pi\omega)^{-\frac{1}{2}}e^{-i\omega
t}\overleftarrow{R}_l(\omega|r)Y_{lm}(\theta,\varphi)\;,
 \eea
where $Y_{lm}(\theta,\varphi)$ are the spherical harmonics and the
radial functions have the following asymptotic forms~\cite{Dewitt75}
 \bea
 \label{asymp1}
&&\overrightarrow{R}_l(\omega|r)\sim\left\{
                    \begin{aligned}
                 &r^{-1}e^{i\omega
r_\ast}+\overrightarrow{A}_l(\omega)r^{-1}e^{-i\omega r_\ast},\;\;r
\rightarrow 2M\;,\cr
                  &
                  {B}_l(\omega)r^{-1}e^{i\omega r_\ast},\;\;\quad\quad \quad\quad \;\;\;\;\;r
\rightarrow\infty\;,\cr
                          \end{aligned} \right.\\
  \label{asymp2}
&&\overleftarrow{R}_l(\omega|r)\sim\left\{
                      \begin{aligned}
&{B}_l(\omega)r^{-1}e^{-i\omega r_\ast},\;\;\quad\quad \quad\quad
\;\;\;r \rightarrow2M\;,\cr &r^{-1}e^{-i\omega
r_\ast}+\overleftarrow{A}_l(\omega)r^{-1}e^{i\omega r_\ast},\;\;r
\rightarrow \infty\;,
                       \end{aligned} \right.
 \eea
with $r_\ast=r+2M\ln(r/2M-1)$ being the Regge-Wheeler tortoise
coordinate. The physical interpretation of these modes is that
$\overrightarrow u$ represents modes emerging from the past horizon
and the $\overleftarrow u$ denotes those coming in from infinity.
With the basics of the scalar field modes given above, we now apply
the formalism outlined in the preceding section to calculate the
Lamb shifts of the static atoms in three vacuum states of the
quantum scalar fields respectively.

\paragraph{Boulware vacuum.}
The Boulware vacuum is defined by requiring normal modes to be
positive frequency with respect to the Killing vector $\partial/
\partial t$. One can  show that the Wightman function for
massless scalar fields in this vacuum state is given
by~\cite{Fulling77,Candelas80}
 \bea
D_B^+(x,x')=\frac{1}{4\pi}\sum_{lm}|Y_{lm}(\theta,\varphi)|^2\,
    \int_{0}^{+\infty}\frac{d\omega}{\omega}
    e^{-i\omega\Delta t}[\,|\overrightarrow{R}_l(\omega|r)|^2
    +|\overleftarrow{R}_l(\omega|r)|^2]\;
 \eea
and the corresponding symmetric correlation function and linear
susceptibility of the field are respectively
  \bea
C^F(x(\tau),x(\tau'))&=&\frac{1}{8\pi}\sum_{lm}|Y_{lm}(\theta,\varphi)|^2
      \int_{0}^{\infty}\frac{d\omega}{\omega}
      \biggl(e^{\frac{i\omega\Delta\tau}{\sqrt{g_{00}}}}+e^{-\frac{i\omega\Delta\tau}
      {\sqrt{g_{00}}}}\biggr)\times\nonumber\\&&\;\quad\;
      [\,|\overrightarrow{R}_l(\omega|r)|^2
      +|\overleftarrow{R}_l(\omega|r)|^2]\;,
 \eea
 \bea
\chi^F(x(\tau),x(\tau'))&=&\frac{1}{8\pi}\sum_{lm}|Y_{lm}(\theta,\varphi)|^2
      \int_{0}^{\infty}\frac{d\omega}{\omega}
      \biggl(e^{-\frac{i\omega\Delta\tau}{\sqrt{g_{00}}}}-e^{\frac{i\omega\Delta\tau}
      {\sqrt{g_{00}}}}\biggr)\times\nonumber\\&&\;\quad\;
      [\,|\overrightarrow{R}_l(\omega|r)|^2
      +|\overleftarrow{R}_l(\omega|r)|^2]\;,
 \eea
where we have used $\Delta\tau= \sqrt{g_{00}}\Delta t$. Substituting
the above statistical functions into Eqs.~(\ref{Evf}) and
(\ref{Err}), extending the integration range for $\tau$ to infinity
for sufficiently long times $\tau-\tau_0$, and performing the
integration over $\tau$, we obtain the contribution of the vacuum
fluctuations to energy level shift for an atom in state $|b\rangle$
held static at a distance $r$ from the black hole
 \bea
(\delta E_b)_{vf}&=&\frac{\mu^2}{32\pi^2\hbar}\sum_{d}|\langle
b|S_2(0)|d\rangle|^2
          \times\nonumber\\&&\;\quad\;\quad\;\int_{0}^{\infty}d\omega
          \biggl(\frac{\omega}{\frac{\omega}{\sqrt{g_{00}}}+\omega_{bd}}-
          \frac{\omega}{\frac{\omega}{\sqrt{g_{00}}}-\omega_{bd}}\biggr)
          [\overrightarrow{g}(\omega|r)+\overleftarrow{g}(\omega|r)]\;,
          \label{Boulware vf}
 \eea
and that of radiation reaction
 \bea
(\delta
E_b)_{rr}&=&-\frac{\mu^2}{32\pi^2\hbar}\sum_{d}|\langle\,b|S_2(0)|d\rangle|^2
          \times\nonumber\\&&\;\quad\;\quad\;\int_{0}^{\infty}d\omega
          \biggl(\frac{\omega}{\frac{\omega}{\sqrt{g_{00}}}+\omega_{bd}}+
          \frac{\omega}{\frac{\omega}{\sqrt{g_{00}}}-\omega_{bd}}\biggr)
          [\overrightarrow{g}(\omega|r)+\overleftarrow{g}(\omega|r)]\;,
          \label{Boulware rr}
 \eea
where we have appealed to the following property of the spherical
harmonics
 \beq
\sum^l_{m=-l}|\,Y_{lm}(\,\theta,\varphi\,)\,|^2= {2l+1 \over
 4\pi}\;
 \eeq
and defined
  \bea
\overrightarrow{g}(\omega|r)=\frac{1}{\omega^2}
\sum_{l=0}^{\infty}(2l+1)|\overrightarrow{R}_l(\omega|r)|^2\;,\\
\overleftarrow{g}(\omega|r)=\frac{1}{\omega^2}
\sum_{l=0}^{\infty}(2l+1)|\overleftarrow{R}_l(\omega|r)|^2\;.
 \eea

Since we do not have a generic expression for ${R}_l(\omega|r)$
functions, let us now examine in detail the energy level shifts both
at close to the horizon and at infinity, which are anyway regions of
physical interest.
Then, the summations in Eqs.~(\ref{Boulware vf}) and
(\ref{Boulware rr}) can be simplified by the asymptotic properties
of the radial functions~\cite{Candelas80}
 \bea
&&\label{asymp1}
          \sum_{l=0}^\infty(2l+1)|\overrightarrow{R}_l(\omega|r)|^2\sim\left\{
          \begin{aligned}
          &\frac{4\omega^2}{1-\frac{2M}{r}},\quad\;\quad\;\quad\;\quad\;\quad\;\quad
              r\rightarrow2M\;, \cr
          &\frac{1}{r^2}\sum_{l=0}^\infty(2l+1)|{B}_l(\omega)|^2\;,\quad r\rightarrow\infty\;,\cr
          \end{aligned} \right.\\
&&\label{asymp2}
         \sum_{l=0}^\infty(2l+1)|\overleftarrow{R}_l(\omega|r)\,|^2\sim\left\{
         \begin{aligned}
         &\frac{1}{4M^2}\sum_{l=0}^\infty(2l+1)|{B}_l(\omega)|^2\;,r\rightarrow2M\;,\cr
         &\frac{4\omega^2}{1-\frac{2M}{r}}\;,
              \quad\;\quad\;\quad\;\quad\;\quad\quad r\rightarrow\infty\;.
         \end{aligned} \right.
         \label{leftarrowR}
 \eea
In Eq.~(\ref{leftarrowR}), we have retained a factor
$g_{00}^{-1}=(1-2M/r)^{-1}$ for the asymptotic form at infinity
which was omitted in Ref.~\cite{Candelas80} as it approaches to $1$
when $r\rightarrow\infty$.

So, in these two asymptotic regions, we can write
 \beq
(\delta E_b)_{vf}=\frac{\mu^2}{8\pi^2\hbar}\sum_{d}|\langle
           b|S_2(0)|d\rangle|^2
           \int_{0}^{\infty}d\omega\biggl[1
           +f(\omega,r)\biggr]
           P\biggl(\frac{\omega}{\omega+\omega_{bd}}
           -\frac{\omega}{\omega-\omega_{bd}}\biggr)\;,
           \label{Boulware vf asymp}
 \eeq
 \beq
 (\delta
E_b)_{rr}=-\frac{\mu^2}{8\pi^2\hbar}\sum_{d}|\langle
           b|S_2(0)|d\rangle|^2
           \int_{0}^{\infty}d\omega\biggl[1
           +f(\omega,r)\biggr]P
           \biggl(\frac{\omega}{\omega+\omega_{bd}}
           +\frac{\omega}{\omega-\omega_{bd}}\biggr)\;.
           \label{Boulware rr asymp}
 \eeq
Here we have made a variable transformation,
$\omega/\sqrt{g_{00}}\rightarrow\omega$, and defined
 \beq
f(\omega,r)=\frac{1}{4r^2\omega^2}
    \sum_{l=0}^{\infty}(2l+1)|B_l(\omega\sqrt{g_{00}})|^2\;,
 \label{grey-body factor}
 \eeq
$P$ here and after denotes the principal value. Notice that for each
definite initial state of a two-level atom, $|b\rangle$,
$\sum_{d}|\langle b|S_2(0)|d\rangle|^2=1/4$, and the integrand in
Eq.~(\ref{Boulware rr asymp}) is an even function of
$|\omega_{bd}|=\omega_0$, so it is obvious that $(\delta
E_+)_{rr}=(\delta E_-)_{rr}$. Radiation reaction has equal
contribution to the energy shift of each level. Adding up the
contributions of vacuum fluctuations and radiation reaction, we
obtain the total energy level shift of the state $|b\rangle$,
 \beq
\delta E_b=-\frac{\mu^2}{4\pi^2\hbar}\sum_{d}|\langle
           b|S_2(0)|d\rangle|^2\int_{0}^{\infty}d\omega
           \biggl[1+f(\omega,r)\biggr]
           P\frac{\omega}{\omega-\omega_{bd}}\;.
 \eeq
So, the energy shift of the excited and ground states of the
two-level atom are respectively
 \bea
 \delta
E_+=-\frac{\mu^2}{16\pi^2\hbar}
           \int_{0}^{\infty}d\omega\biggl[1+f(\omega,r)\biggr]
           P\frac{\omega}{\omega-\omega_{0}}
           \;,\\
\delta E_-=-\frac{\mu^2}{16\pi^2\hbar}\int_{0}^{\infty}d\omega
           \biggl[1+f(\omega,r)\biggr]
           P\frac{\omega}{\omega+\omega_{0}}\;.
 \eea
The relative energy shift, i.e., the Lamb shift, which is an
observable physical quantity, is then obtained by the subtraction,
$\Delta= \delta E_+- \delta E_-$
  \bea
\Delta_B&=&\frac{\mu^2}{16\pi^2\hbar}\int_0^\infty d\omega
        P\biggl(\frac{\omega}{\omega+\omega_0}
        -\frac{\omega}{\omega-\omega_0}\biggr)\nonumber\\&
        +&\frac{\mu^2}{16\pi^2\hbar}
       \int_0^\infty d\omega f(\omega,r) P\biggl(\frac{\omega}{\omega+\omega_0}
        -\frac{\omega}{\omega-\omega_0}\biggr)\;.
        \label{DeltaB two part}
 \eea
Actually, the relative shift of the atomic energy is entirely caused
by vacuum fluctuations and can be calculated directly by
$\Delta=(\delta E_+)_{vf}-(\delta E_-)_{vf}$, because of the equal
contribution of radiation reaction to the two levels. It is composed
of two parts. The first part is just the Lamb shift of a two-level
atom in a free Minkowski space-time with no boundaries. It is
logarithmically divergent, which is expected for a nonrelativistic
treatment as what we do here and the divergence can be removed by
introducing a cutoff~\cite{H.A.Bethe47,Welton48} or resorting to a
fully relativistic approach~\cite{French-Weisskopf49,Kroll-Lamb49}.
The second part represents a finite correction to the Lamb shift in
an unbounded flat space-time. It arises as a result of the back
scattering of vacuum field modes off the space-time curvature of the
black hole in much the same way as the reflection of the field modes
at the reflecting boundary in a flat space-time, which also gives
rise to corrections to the Lamb shift in the unbounded
space~\cite{Meschede90,L.Rizzuto07}. In fact, the second part of
Eq.~(\ref{DeltaB two part}) gives the correction to the Lamb shift
for an atom held static outside a massive spherical object with a
radius larger than the Schwarzschild radius, since the Boulware
vacuum is the natural vacuum outside such an object. This part can
actually be further simplified. To do this, let us note that  the
coefficient $|B_l|$ can be approximated, using the geometrical
optics approximation\cite{Dewitt75},  by 1 for $l <\sqrt{27}
M\omega$ and 0 for $l >\sqrt{27} M\omega$ so that
$B_l(\omega)\sim\theta(\sqrt{27}M\omega-l)$. Thus, the summation in
$f(\omega,r)$ can be evaluated to yield
 \bea
f(\omega,r)
\approx\frac{27M^2g_{00}}{4r^2}=\frac{27M^2}{4r^2}\bigg(1-{2M\over
r}\bigg)\equiv f(r)\;,
 \eea
which does not depend on $\omega$. So, the second part is also
logarithmically divergent and the divergence can again be dealt with
as that in the first part. Denote the Lamb shift in a free Minkowski
space-time by $\Delta_M$, we can write
 \bea
\Delta_B=[1+f(r)]\;\Delta_M \;.
 \eea
Let us note that the correction vanishes both at infinity and at the
even horizon,  where the effective potential becomes zero,  and it
reaches the maximum value near $r=3M$, where the effective potential
has a peak and thus the vacuum field modes are most strongly
scattered.

\paragraph{Unruh vacuum.}
Now we turn our attention to the case of the Unruh vacuum. The
Wightman function for the real massless scalar field in this vacuum
is given by~\cite{Fulling77,Candelas80}
 \bea
D_U^+(x,x')&=&\frac{1}{4\pi}\sum_{lm}|Y_{lm}(\theta,\varphi)|^2
    \int_{-\infty}^{+\infty}\frac{d\omega}{\omega}\times\nonumber\\&&
    \quad\;\biggl[\frac{e^{-i\omega\Delta t}}{1-e^{-2\pi\omega/\kappa}}
    |\overrightarrow{R}_l(\omega|r)|^2
    +\theta(\omega)e^{-i\omega\Delta
    t}|\overleftarrow{R}_l(\omega|r)|^2\biggr]\;,
 \eea
where $\kappa=1/4M$ is the surface gravity of the black hole. The
two statistical functions of the scalar field readily follow
 \bea
C^F(x(\tau),x(\tau'))&=&\frac{1}{8\pi}\sum_{lm}|Y_{lm}(\theta,\varphi)|^2
     \int_{-\infty}^{\infty}\frac{d\omega}{\omega}
     \biggl(e^{\frac{i\omega\Delta\tau}{\sqrt{g_{00}}}}+
     e^{-\frac{i\omega\Delta\tau}{\sqrt{g_{00}}}}\biggr)\times\nonumber\\&&\quad\;
     \biggl[\frac{|\overrightarrow{R}_l(\omega|r)|^2}{1-e^{-2\pi\omega/\kappa}}
     +\theta(\omega)|\overleftarrow{R}_l(\omega|r)|^2\biggr]\;,
 \eea
 \bea
\chi^F(x(\tau),x(\tau'))&=&\frac{1}{8\pi}\sum_{lm}|Y_{lm}(\theta,\varphi)|^2
     \int_{-\infty}^{\infty}\frac{d\omega}{\omega}
     \biggl(e^{-\frac{i\omega\Delta\tau}{\sqrt{g_{00}}}}-e^{\frac{i\omega\Delta\tau}
     {\sqrt{g_{00}}}}\biggr)\times\nonumber\\&&\quad\;
     \biggl[\frac{|\overrightarrow{R}_l(\omega|r)|^2}{1-e^{-2\pi\omega/\kappa}}
     +\theta(\omega)|\overleftarrow{R}_l(\omega|r)|^2\biggr]\;.
 \eea
The contributions of vacuum fluctuations and radiation reaction can
now be calculated by Eqs.~(\ref{Evf}) and (\ref{Err}) to yield
 \bea
(\delta E_b)_{vf}&=&\frac{\mu^2}{32\pi^2\hbar}\sum_{d}|\langle
          b|S_2(0)|d\rangle|^2\int_{-\infty}^{\infty}d\omega\times\nonumber\\&&
          \quad\;\quad\;\biggl[\frac{1}{1-e^{-2\pi\omega/\kappa}}\overrightarrow{g}(\omega|r)
          +\theta(\omega)\overleftarrow{g}(\omega|r)\biggr]
          \biggl(\frac{\omega}{\frac{\omega}{\sqrt{g_{00}}}+\omega_{bd}}
          -\frac{\omega}{\frac{\omega}{\sqrt{g_{00}}}-\omega_{bd}}\biggr)
          \;,
 \eea
 \bea
(\delta E_b)_{rr}&=&-\frac{\mu^2}{32\pi^2\hbar}\sum_{d}|\langle
          b|S_2(0)|d\rangle|^2\int_{-\infty}^{\infty}d\omega
          \times\nonumber\\&&\quad\;\quad\;
          \biggl[\frac{1}{1-e^{-2\pi\omega/\kappa}}\overrightarrow{g}(\omega|r)
          +\theta(\omega)\overleftarrow{g}(\omega|r)\biggr]
          \biggl(\frac{\omega}{\frac{\omega}{\sqrt{g_{00}}}+\omega_{bd}}
          +\frac{\omega}{\frac{\omega}{\sqrt{g_{00}}}-\omega_{bd}}\biggr)
          \;.\label{Unruh general rr}
 \eea
Here again, one can see from Eq.(\ref{Unruh general rr}) that
$(\delta E_+)_{rr}=(\delta E_-)_{rr}$ for the same reasons as that
in the Boulware vacuum. Similarly, we now focus our attention on the
two asymptotic regions, i.e., when $r \rightarrow2M$ and $r
\rightarrow\infty$. Adding up the contributions of vacuum
fluctuations and radiation reaction, calculating out the energy
shift for each level, and then performing a subtraction, we can
derive the Lamb shift, which is, for an atom fixed near the event
horizon, i.e., when $r\rightarrow2M$,
 \beq
\Delta_U\approx[1+f(r)]\;\Delta_M +\Delta_T\;,
      \label{DeltaU2M}
 \eeq
 where
 \beq
\Delta_T=\frac{\mu^2}{8\pi^2\hbar}\int_{0}^{\infty}d\omega
           P\biggl(\frac{\omega}{\omega+\omega_{0}}
           -\frac{\omega}{\omega-\omega_{0}}\biggr)
           \frac{1}{e^{2\pi\omega/\kappa_r}-1}\;,
 \eeq
and $\kappa_r=\kappa/\sqrt{g_{00}}$. $\Delta_T$, the correction
term, as opposed to the Boulware vacuum case, is in the same form as
the acceleration correction to the Lamb shift derived in
\cite{Audretsch95} of a uniformly accelerated atom which would find
itself in a thermal bath at the Unruh temperature, and is in
structural similarity to the corresponding expressions obtained for
the Lamb shift in a thermal heat bath~\cite{Barton72,Knight72,Farley
and Wing81} in that they have in common the appearance of the
thermal factor $(e^{2\pi\omega/\kappa_r}-1)^{-1}$. The difference
can be attributed to the discrepancy between the scalar field we
consider here and the electromagnetic field.

So, close to the horizon, the Lamb shift gets a correction, which is
what one would find for a static atom if there is thermal radiation
at the temperature
  \beq
T=\frac{\kappa_r}{2\pi}=\frac{\kappa}{2\pi\sqrt{g_{00}}}=\frac{T_H}{\sqrt{g_{00}}}\;
\label{tem-r}
 \eeq
with $T_H=\kappa/2\pi$, being the usual Hawking temperature of the
black hole. Eq.~({\ref{tem-r}) is the well-known Tolman
relation~\cite{Tolman304,Tolman301}, which gives the proper
temperature as measured by a local observer. Thus, a static atom
close to the horizon of a black hole behaves, in terms of the Lamb
shift, as if there is thermal radiation emanating from the black
hole horizon. Notice that $T$ actually diverges as the horizon is
approached, and this should come as no surprise since the atom must
be in acceleration relative to the local free-falling frame to
maintain at a fixed distance from the black hole, and this
acceleration, which blows up at the horizon, gives rise to
additional thermal effect. A comparison of Eq.~(\ref{DeltaU2M}) with
Eq.(\ref{DeltaB two part}) leads to the following relationship at
the horizon
 \beq
\Delta_U|_{\;r\rightarrow2M}=\Delta_B|_{\;r\rightarrow2M}
+\;\Delta_T|_{\;r\rightarrow2M}\;.
 \eeq
For an atom fixed far from the black hole, i.e. when
$r\rightarrow\infty$, the Lamb shift becomes
 \beq
\Delta_U\approx[1+f(r)]\;\Delta_M + f(r)\Delta_T\;.
 \label{Unruhinfinity}
 \eeq
Here we have used the relation $f(-\omega,r)=f(\omega,r)$ that can
be deduced from the properties of $B_l(\omega)$ given in
Ref.~\cite{Dewitt75}. Different from that near the event horizon,
the correction term, which is thermallike, is modified by a
grey-body factor $f(\omega,r)\sim f(r)$. It can be understood as a
result of backscattering of the outgoing thermal flux emanating from
the event horizon off space-time curvature, which results in part of
the outgoing flux being depleted. As the atom is placed further and
further away, the thermal flux becomes weaker and weaker, and so is
its contribution to the Lamb shift. Consequently,
  \bea
\Delta_U|_{r\;\rightarrow\infty}&\approx&\Delta_B|_{\;r\rightarrow\infty}
        +\;[f(r)\Delta_T]_{\;r\;\rightarrow\infty}\nonumber\\
        &\approx&\Delta_B|_{\;r\rightarrow\infty}=\Delta_M\;.
        \label{deltaE two part}
 \eea

\paragraph{Hartle-Hawking vacuum.}
Let us focus on the case of the Hartle-Hawking vacuum. The Wightman
function for the real massless scalar field now
is~\cite{Fulling77,Candelas80}
 \bea
D_H^+(x,x')&=&\frac{1}{4\pi}\sum_{lm}|Y_{lm}(\theta,\varphi)|^2\
    \int_{-\infty}^{+\infty}\frac{d\omega}{\omega}\times
    \nonumber\\&&\quad\;\biggl[\frac{e^{-{i\omega\Delta t}
    }}{1-e^{-2\pi\omega/\kappa}}|\overrightarrow{R}_l(\omega|r)|^2
    +\frac{e^{{i\omega\Delta t}}}{e^{2\pi\omega/\kappa}-1}
    |\overleftarrow{R}_l(\omega|r)|^2\biggr]\;,
 \eea
and it leads to the statistical functions of the scalar field as
follows
 \bea
C^F(x(\tau),x(\tau'))&=&\frac{1}{8\pi}\sum_{lm}|Y_{lm}(\theta,\varphi)|^2
   \int_{-\infty}^{+\infty}\frac{d\omega}{\omega}\biggl(e^{\frac{i\omega\Delta\tau}
   {\sqrt{g_{00}}}}+e^{-\frac{i\omega\Delta\tau}{\sqrt{g_{00}}}}\biggr)
   \nonumber\\&&\quad\;\times
   \biggl(\frac{|\overrightarrow{R}_l(\omega|r)|^2}{{1-e^{-{2\pi\omega}/{\kappa}}}}
   +\frac{|\overleftarrow{R}_l(\omega|r)|^2}{{e^{{2\pi\omega}/{\kappa}}-1}}\biggr)\;,
 \eea
 \bea
\chi^F(x(\tau),x(\tau'))&=&\frac{1}{8\pi}\sum_{lm}|Y_{lm}(\theta,\varphi)|^2
   \int_{-\infty}^{+\infty}\frac{d\omega}{\omega}\biggl(e^{-\frac{i\omega\Delta\tau}
   {\sqrt{g_{00}}}}-e^{\frac{i\omega\Delta\tau}{\sqrt{g_{00}}}}\biggr)
   \nonumber\\&&\quad\;\times
   \biggl(\frac{|\overrightarrow{R}_l(\omega|r)|^2}{{1-e^{-{2\pi\omega}/{\kappa}}}}
   -\frac{|\overleftarrow{R}_l(\omega|r)|^2}{{e^{{2\pi\omega}/{\kappa}}-1}}\biggr)\;.
 \eea
So, the contributions of vacuum fluctuations and radiation reaction
to the energy shift of level $|b\rangle$ can be  separately
calculated to give
 \bea
(\delta E_b)_{vf}&=&\frac{\mu^2}{32\pi^2\hbar}\sum_{d}|\langle
b|S_2(0)|d\rangle|^2\int_{-\infty}^{\infty}d\omega\times
          \nonumber\\&&\quad\;\quad
          \biggl(\frac{\omega}{\frac{\omega}{\sqrt{g_{00}}}+\omega_{bd}}
          -\frac{\omega}{\frac{\omega}{\sqrt{g_{00}}}-\omega_{bd}}\biggr)
          \biggl[\frac{\overrightarrow{g}(\omega|r)}{1-e^{-2\pi\omega/\kappa}}
          +\frac{\overleftarrow{g}(\omega|r)}{e^{2\pi\omega/\kappa}-1}\biggr]\;,
 \eea
 \bea
(\delta E_b)_{rr}&=&-\frac{\mu^2}{32\pi^2\hbar}\sum_{d}|\langle
b|S_2(0)|d\rangle|^2\int_{-\infty}^{\infty}d\omega\times
          \nonumber\\&&\quad\;\quad\;\quad
          \biggl(\frac{\omega}{\frac{\omega}{\sqrt{g_{00}}}+\omega_{bd}}
          +\frac{\omega}{\frac{\omega}{\sqrt{g_{00}}}-\omega_{bd}}\biggr)
          \biggl[\frac{\overrightarrow{g}(\omega|r)}{1-e^{-2\pi\omega/\kappa}}
          -\frac{\overleftarrow{g}(\omega|r)}{e^{2\pi\omega/\kappa}-1}\biggr]\;.
          \label{Hartle rr}
 \eea
Using the above results, we can examine the behaviors of the Lamb
shift in two asymptotic regions. First, when $r\rightarrow\infty$,
we find
 \beq
\Delta_H\approx[1+f(r)]\;\Delta_M + f(r)\;\Delta_T+\;\Delta_T\;.
         \label{DeltaHinfty}
 \eeq
Here, the first correction term as opposed to the Boulware vacuum
case is caused by the thermal radiation from the black hole which is
backscattered by the space-time curvature and the backscattering is
represented by the appearance of the grey-body factor, $f(\omega,
r)\sim f(r)$, as we have already studied in the Unruh vacuum case.
The second correction term ($\Delta_T$) is the correction  one would
get for the Lamb shift in a thermal bath at the Hawking temperature
$T_H$ (Here we have taken into account that $T=T_H$ at infinity.)
Since the thermal radiation from the black hole diminishes to zero
at infinity due to the backscattering off the curvature, as we have
already pointed out, the presence of the thermal correction term,
$\Delta_T$, indicates, in the Hartle-Hawking vacuum, that there is a
thermal distribution of quanta at the Hawking temperature at
infinity, and therefore, the Hartle-Hawking vacuum is not a vacuum
in real sense, but a state that describes a black hole in
equilibrium with an infinite sea of blackbody radiation. This is
consistent with our understanding of the Hartle-Hawking vacuum
gained in studies in other different
contexts~\cite{Candelas80,hyprd07,ZhangYu07}. Note that
Eq.~(\ref{DeltaHinfty}) can also be written as
 \beq
\Delta_H|_{\;r\rightarrow\infty}=\;\Delta_U|_{\;r\rightarrow\infty}\;
+\;\Delta_T|_{\;r\rightarrow\infty}\;.
 \eeq

Let us now study what happens at close to the horizon, i.e., when
$r\rightarrow2M$. It then follows that
 \beq
\Delta_H\approx[1+f(r)]\;\Delta_M + [1+f(r)]\;\Delta_T\;.
 \label{DeltaH2M}
 \eeq
Here the $\Delta_T$ term is a correction reminiscent of what one has
for the Lamb shift close to the black hole horizon in the Unruh
vacuum case, and it is a consequence of the outgoing thermal
radiation from the black hole. The last term is a contribution of
the incoming radiation of the thermal bath at infinity, which is
backscattered off the curvature on its way to the black hole
horizon. Eq.~(\ref{DeltaH2M}) can also be written as
 \bea
\Delta_H|_{\;r\rightarrow2M}=\;\Delta_U|_{\;r\rightarrow2M}
        +\;[f(r)\Delta_T]_{\;r\rightarrow2M}\;.
 \eea
It is interesting to note that the appearance of the $\Delta_T$ term
in the Lamb shifts close to the horizon both in the Unruh and the
Hartle-Hawking vacua (refer to Eq.~(\ref{DeltaU2M}) and
Eq.~(\ref{DeltaH2M})) supports the notion that the Hawking radiation
of a black hole originates from the black hole horizon, since
$\Delta_T$ represents a contribution of purely thermal radiation not
scattered by the space-time curvature.

Finally, it is worth pointing out that the correction term
$\Delta_T$, which appears in the Lamb shifts in both the Unruh and
Hartle-Hawking vacuua, is finite, and this can be seen by a careful
inspection of the integral involved. Here, we would like to go a
little bit further. We will analyze the behaviors of $\Delta_T$ both
in the high and low temperature limits. For this purpose, we rewrite
it as
 \bea
\Delta_T=-\frac{\mu^2\omega_0}{8\pi^2\hbar}\int_{0}^{\infty}dx
           P\biggl(\frac{1}{x+x_0}+\frac{1}{x-x_0}\biggr)
           \frac{1}{e^{x}-1}\;
 \eea
with $x=\omega/T$ and $x_0=\omega_0/T$. When the Hawking temperature
is low, that is, when $x_0\gg1$, the integral can be approximated by
 \beq
\Delta_T=\frac{\mu^2}{8\hbar}
   \biggl[\frac{T^2}{3 \omega_0}+\frac{2\pi^2 T^4}{15 \omega_0^3}
   +O\biggl(\frac{1}{x_0^6}\biggr)\biggr]\;.
 \eeq
While in the high temperature limit, i.e., when $x_0\ll1$, we have
 \beq
\Delta_T\approx\frac{\mu^2\omega_0}{8\pi^2\hbar}\frac{1}{x_0}
         \biggl[e^{-x_0}\overline{Ei}(x_0)
         -e^{x_0}\overline{Ei}(-x_0)\biggr]\;,
 \eeq
where $Ei(z)=-\int_{-z}^\infty\frac{e^{-t}}{t}dt$ and the overline
 denotes the principal value. Further simplifications yield
 \bea
\Delta_T\approx\frac{\mu^2\omega_0}{4\pi^2\hbar}
        [1-\gamma+\ln (T/\omega_0)]+O(x_0^2)\;
 \eea
with $\gamma=0.577216$ being the Euler's constant. So, in the high
temperature regime, the Lamb shift increases logarithmically with
the temperature.

\section{Summary}
Using the formalism suggested by Dalibard, Dupont-Roc and
Cohen-Tannoudji(DDC)~\cite{Dalibard82,Dalibard84} which allows a
distinct separation of the contributions of vacuum fluctuations and
radiation reaction, we have calculated the Lamb shift of a static
two-level atom interacting with real massless scalar fields in the
Boulware, Unruh and Hartle-Hawking vacuums outside a Schwarzschild
black hole.

In the Boulware vacuum case, we find that the Lamb shift gets a
correction, as opposed to that in an unbounded flat space,  which is
reminiscent of the correction to the Lamb shift induced by the
presence of boundaries, and it can be understood as a result of the
back scattering of vacuum field modes off the space-time curvature
of the black hole in much the same way as the reflection of the
field modes at the reflecting boundary in a flat space-time. This
correction can be viewed as the correction to the Lamb shift for an
atom held static outside a massive spherical object with a radius
larger than the Schwarzschild radius, and it reduces,  at infinity,
to the Lamb shift in a unbounded flat space as expected.

In the Unruh vacuum case, we find that the Lamb shift is corrected
by a thermal term, as compared to that in Boulware vacuum case,
which can be regarded as a result of the thermal radiation at the
Hawking temperature  emanating from the black hole horizon and the
additional thermal effect due to the acceleration  relative to the
local free-falling frame  the atom must have in order to be static.
However, the thermal radiation from the black hole is backscattered
off the space-time curvature, and becomes weaker and weaker on its
way away from the black hole, rendering the Lamb shift to reduce to
that in a flat space at infinity.

However, when the Hartle-Hawking vacuum is concerned,  our results
show that the correction to the Lamb shift close to the horizon, as
opposed to that in the Boulware vacuum, would be a purely thermal
contribution term due to an outgoing Hawking radiation from black
hole plus that due to  an incoming radiation from infinity which is
backscattered off the curvature, whereas, at infinity, the Lamb
shift gets corrected by a purely thermal term characteristic of a
sea of black-body radiation plus a term resulting from the
backscattered Hawking radiation from the black hole.  This supports
the notion that the Hartle-Hawking vacuum is  a state that describes
a black hole in equilibrium with an infinite sea of blackbody
radiation, rather than a vacuum state in the usual sense.

\begin{acknowledgments}

One of us (WZ) would like to thank Jialin Zhang for helpful
discussions. This work was supported in part by the National Natural
Science Foundation of China under Grants No. 10775050, No. 11075083,
and No. 10935013; the Zhejiang Provincial Natural Science Foundation
of China under Grant No. Z6100077; the SRFDP under Grant No.
20070542002; the National Basic Research Program of China under
Grant No. 2010CB832803; the PCSIRT under Grant No. IRT0964; and the
Programme for the Key Discipline in Hunan Province.

\end{acknowledgments}


\begin{thebibliography}{90}

\bibitem{Meschede90}D. Meschede, W. Jhe and E.A. Hinds, Phys. Rev. A {\bf 41}, 1587 (1990).
\bibitem{Audretsch95} J. Audretsch and R. M\"uller,
Phys. Rev. A {\bf 52}, 629 (1995).
\bibitem{Passante98} R. Passante, Phys. Rev. A {\bf 57}, 1590 (1998).
\bibitem{L.Rizzuto07} L. Rizzuto, Phys. Rev. A {\bf 76}, 062114 (2007).
\bibitem{ZhuYu10}Z. Zhu and H. Yu, Phys. Rev. A, (to be published).
\bibitem{Unruh} W. G. Unruh, Phys. Rev. D {\bf 14}, 870 (1976).
\bibitem{Hartle-Hawking} J.~Hartle and S.~Hawking, Phys. Rev. {\bf
D13}, 2188 (1976).
\bibitem{Dalibard82} J. Dalibard, J. Dupont-Roc and C. Cohen-Tannoudji, J.
Phys. (France){\bf43}, 1617 (1982).
\bibitem{Dalibard84} J. Dalibard, J. Dupont-Roc and C. Cohen-Tannoudji, J.
Phys. (France){\bf45}, 637 (1984).
\bibitem{hyprd07} H. Yu and W. Zhou, Phys. Rev. D, {\bf
76}, 027503 (2007); {\it ibid}, {\bf 76}, 044023 (2007).
\bibitem{Audretsch94} J. Audretsch and R. M\"uller, Phys. Rev. A {\bf 50},
1755 (1994).
\bibitem{Dewitt75} B. S. DeWitt, Phys. Rep. {\bf 19}, 295 (1975).
\bibitem{Fulling77} S. M. Christensen and S. A. Fulling, Phys. Rev. D {\bf15}, 2088
(1977).
\bibitem{Candelas80} P. Candelas, Phys. Rev. D {\bf21}, 2185 (1980).
\bibitem{H.A.Bethe47} H. A. Bethe, Phys. Rev. {\bf 72}, 339 (1947).
\bibitem{Welton48} T. A. Welton, Phys. Rev. {\bf 74}, 1157 (1948).
\bibitem{Kroll-Lamb49} N. M. Kroll and W. E. Lamb, Phys. Rev. {\bf
75}, 388 (1949).
\bibitem{French-Weisskopf49} J. B. French and V. F.  Weisskopf, Phys. Rev. {\bf
75}, 1240 (1949).
\bibitem{Barton72} G. Barton, Phys. Rev. A {\bf 5}, 468 (1972).
\bibitem{Knight72} P. L. Knight, J. Phys. A {\bf 5}, 417 (1972).
\bibitem{Farley and Wing81} J. W. Farley and W. H. Wing, Phys. Rev. A {\bf 23}, 2397 (1981).
\bibitem{Tolman304} R. Tolman, Phys. Rev. {\bf 35}, 904 (1930).
\bibitem{Tolman301} R. Tolman and P. Ehrenfest, Phys. Rev. {\bf 36}, 1791 (1930).
\bibitem{ZhangYu07} J. Zhang and H. Yu, Phys. Rev. D {\bf 75}, 104014 (2007).


\end{thebibliography}
\end{document}